\documentclass[11pt,acmsmall]{acmart}

\acmConference[Intel Labs]{Intel Labs}{\date{}}{Hillsboro, OR, U.S.}

\renewcommand\footnotetextcopyrightpermission[1]{} 
\makeatletter
\renewcommand\@formatdoi[1]{\ignorespaces}
\makeatother

\setcopyright{none}

\usepackage{booktabs}   
\usepackage{subcaption} 

\usepackage{graphicx}

\usepackage{color}
\usepackage{blindtext}
\usepackage{tcolorbox}
\usepackage{amsmath}
\usepackage{amssymb}
\usepackage{amsthm}
\usepackage{multicol}
\usepackage[inline]{enumitem}

\usepackage{hyperref}
\newcommand{\customurl}[1]{\textcolor{blue}{\textsc{\url{#1}}}}
\newcommand{\refsec}[1]{{Section~\ref{#1}}}
\newcommand{\reffig}[1]{{Fig.~\ref{#1}}}

\newcommand{\pdo}[0]{{PDO}}
\newcommand{\pdos}[0]{{\pdo{}s}}
\newcommand{\scheme}[0]{{Scheme}}
\newcommand{\tinyscheme}[0]{{Tiny\scheme{}}}
\newcommand{\distributedledger}[0]{{DL}}
\newcommand{\trustedexecutionenvironment}[0]{{TEE}}
\newcommand{\provisioningservice}[0]{{PS}}
\newcommand{\enclaveservice}[0]{{ES}}
\newcommand{\smartcontract}[0]{{SC}}
\newcommand{\contract}[0]{{\mathcal{C}}}
\newcommand{\contractcode}[0]{{\mathcal{C}}}
\newcommand{\enclave}[0]{{\mathcal{E}}}
\newcommand{\contractenclave}[0]{{\mathcal{E}_\contractinterpreter{}}}
\newcommand{\contractstate}[0]{{\mathcal{S}_\contract{}}}
\newcommand{\contractowner}[0]{{\mathcal{O}_\contract{}}}
\newcommand{\contractinterpreter}[0]{{\mathcal{I}}}
\newcommand{\ias}[0]{{IAS}}
\newcommand{\partitle}[1]{{#1}}
\begin{document}

\title[PDO]{Private Data Objects: an Overview}        

\author{Mic Bowman, Andrea Miele, Michael Steiner, Bruno Vavala}

\email{firstname.lastname@intel.com}          
\affiliation{
  \department{SPR}              
  \institution{Intel Labs}            
  \country{U.S.}                    
}

\begin{abstract}
We present Private Data Objects (\pdos{}), a technology that enables mutually untrusted parties to run smart contracts over private data. 
\pdos{} result from the integration of a distributed ledger and Intel Software Guard Extensions (SGX). 
In particular, contracts run off-ledger in secure enclaves using Intel SGX, which preserves data confidentiality, execution integrity and enforces data access policies (as opposed to raw data  access).
A distributed ledger verifies and records transactions produced by \pdos{}, in order to provide a single authoritative instance of such objects.
This allows contracting parties to retrieve and check data related to contract and enclave instances, as well as to serialize and commit contract state updates.
The design and the development of \pdos{} is an ongoing research effort, and open source code is available and hosted by Hyperledger Labs~\cite{Hyperledgerlabs2018, Privatedataobjectrepo2018}.

\end{abstract}

\maketitle
\thispagestyle{empty}
\section{Introduction}
\label{sec:introduction}

\partitle{As the community makes progress on the development of distributed ledger (\distributedledger{}) and smart contract (\smartcontract{}) execution architectures, 
the need for verifiability raises privacy and confidentiality concerns.} 
On one hand, the participants in a \distributedledger{} need all the data involved in a transaction in order to perform the appropriate checks and to ensure that it is valid. 
In the case of \smartcontract{}s, this includes having access to the input and output data for executing and validating a \smartcontract{}. 
On the other hand, this requirement clearly clashes with the sensitive nature of information such as, for example, personal data, health records, or confidential corporate documents. 
As a consequence, this problem ends up restricting the application domain of \smartcontract{}s. 

\partitle{The solutions that have been proposed to address this issue range from private groups to cryptographically-secure computation.} 
Private groups can achieve good performance by simply implementing access control mechanisms for group members, under the assumption that such members are trusted.
The latter instead uses complex zero-knowledge constructions or multi-party computation protocols, 
which are based on cryptographic assumptions and are relatively expensive in practice.

\begin{figure}
\includegraphics[width=0.5\linewidth]{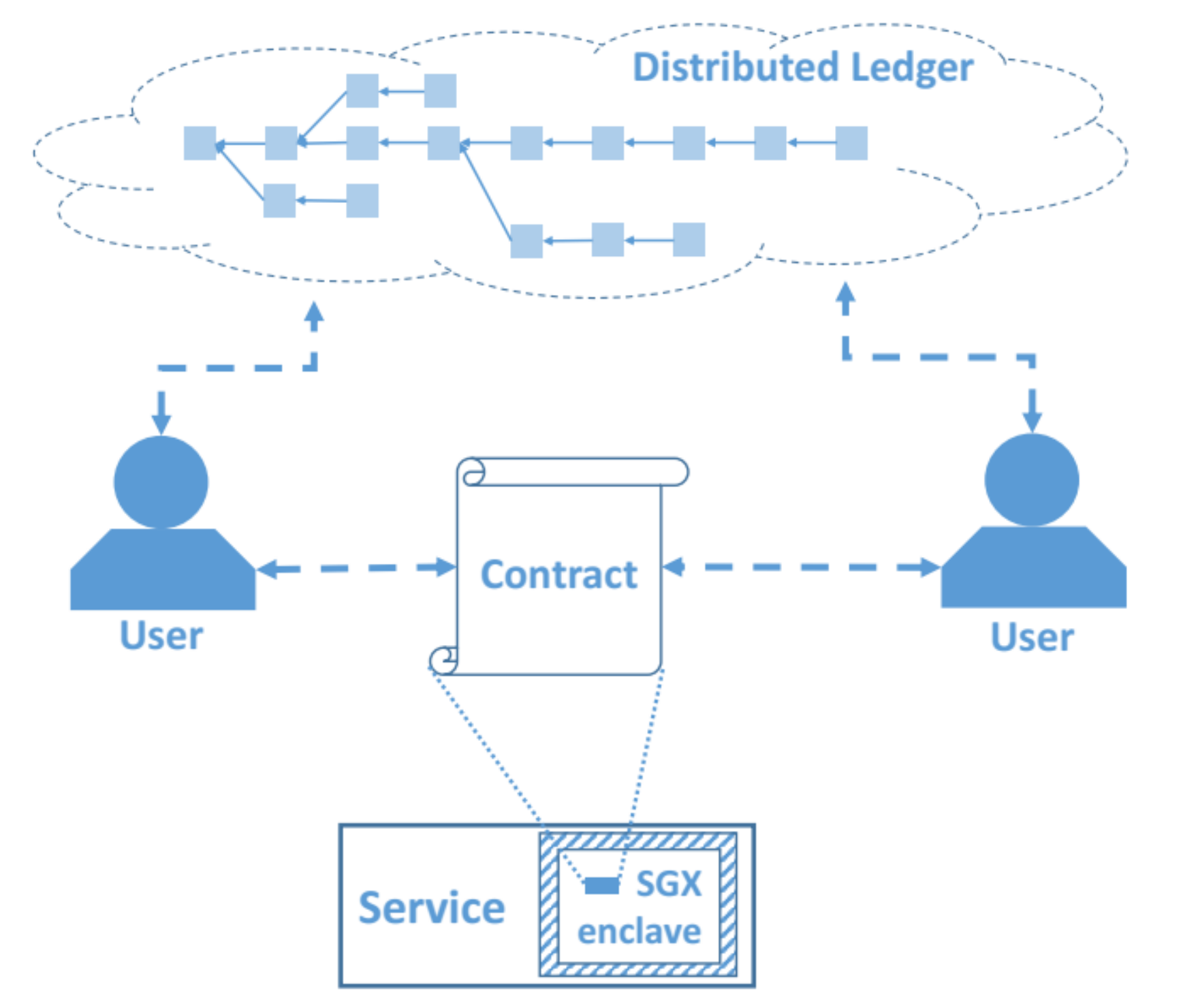}
\caption{\pdo{} concept for private off-chain contract execution.}
\label{fig:pdoconcept}
\end{figure}

\partitle{In this paper we introduce Private Data Objects (\pdo{}s), a solution based on a trusted execution environment (\trustedexecutionenvironment{}) and a \distributedledger{}~(\reffig{fig:pdoconcept}).} 
The \trustedexecutionenvironment{} allows us to run contract code in isolation (from OS and other applications), ensuring execution integrity and data confidentiality. 
The \distributedledger{} allows us to track the registered contracts and enclaves, to serialize the state transitions (or updates) of contracts and to coordinate their interactions. 

\partitle{The combination of these technologies allows \pdos{} to achieve the desired balance between security, performance, efficiency and programmability.} 
More precisely, \pdo{}s leverage Intel SGX to set up a \trustedexecutionenvironment{} for running contract code. 
Since SGX provides execution verifiability, contracts can run off-ledger (on SGX-capable machines). 
This improves efficiency by avoiding the execution of a contract code by possibly numerous mutually-untrusted ledger nodes.
In addition, PDO contracts can implement arbitrary programs, without therefore limiting the application domain.

\partitle{Applications of \pdo{}s are meant to enable mutually-distrustful individuals and organizations to share and process data according to pre-agreed policies, which are defined in the contract itself.}
Such policies to access and update data are carried with the object and enforced by trusted hardware. 
This holds regardless of where the object resides or how often it changes hands. 
However, although in principle any SGX platform can deliver such property, users only accept transactions that are recorded and thus validated on the ledger. 
The responsibility of submitting a transaction (on the ledger) is on the party who initiates the operation resulting in that transaction. \pdos{} let parties use pseudonyms for this task in order to guarantee privacy and unlinkability on the ledger.

\partitle{The rest of this paper is organized as follows.} 
We briefly outline some related projects for privacy preservation in \distributedledger{}s and \smartcontract{}s in \refsec{sec:relatedwork}. 
We specify a list of security threats that guides our design efforts in \refsec{sec:model}. 
Then we introduce the main components of \pdo{}s and explain how they interact with each other in \refsec{sec:overview}. 
We elaborate on the security aspects of \pdo{}s and on the consequences of hardware compromise in \refsec{sec:securityanalysis}. 
Finally, we summarize our findings and conclusions in \refsec{sec:conclusions}.

\section{Related Work}
\label{sec:relatedwork}

Bitcoin and Ethereum are arguably the most popular distributed systems for implementing and executing \smartcontract{}s today.
Also, they provide the underlying distributed ledger.
However, they were not designed to ensure data confidentiality, since the information is public on the ledger.

\paragraph{Privacy and confidentiality in distributed ledgers.}
Several solutions have been proposed to solve the problem of information leakage in distributed ledgers.
Zerocash~\cite{Sasson2014} and Hawk~\cite{Kosba2016} are of particular interest since they provide strong transactional and user privacy as well as programmability.
They both rely on cryptographic primitives (Zero-Knowledge arguments).
Hawk additionally relies on a facilitating party which can be instantiated with Trusted Computing technology (e.g., Intel SGX).

Enigma~\cite{Zyskind2015} proposes the use of Secure Multi-Party Computation as a building block for the decentralized execution of \smartcontract{}s on private data, while relying on an external blockchain-based \distributedledger{} for identity management, access control and a tamper-proof event log.
As this may require the action of several nodes, Enigma reduces redundancy by replicating computation and storage to a subset of the participating nodes.

Solidus~\cite{Cecchetti2017} achieves transaction-graph confidentiality by enhancing Oblivious RAM with public verifiability.
In particular, the new construction enables clients to update a public (on the ledger, though encrypted) physical memory and to generate a zero-knowledge proof of correct update.

The Global Synchronization Log (GSL)~\cite{DigitalAsset2018} bears similarities with \pdo{}s in that it proposes to separate contract execution and data from the ledger.
This allows GSL to limit the exposure of sensitive data to the entitled participants.
Differently from \pdo{}s, and similarly to Fabric (described next), GSL does not constrain the trusted computing base, nor it enforces the as-intended execution of contracts (i.e., entitled parties can misbehave).

Ekiden~\cite{cheng2018ekiden} provides a platform for \smartcontract{} execution that is close to \pdos{}. 
The main similarities lie in the off-ledger contract execution, the SGX-based data privacy approach, and persisting the contract state on the ledger for availability and consistency. 
Ekiden and \pdos{} however diverge on some key aspects, such as: enclaves are stateless and not contract-specific in \pdos{}, thereby providing a contract execution service on-demand; \pdos{} ecosystem includes key provisioning services which enable enclave selection and migration of contract executions, while Ekiden has a set of key managers that replicate keys on all of the compute nodes.
In general, \pdos{} focuses more on real-world aspects such as scalability -- e.g., expressing dependencies via CCL allows
more asynchronous in execution of interacting contracts than enforcing
them post-facto via proof of publications -- while at the same time
providing secondary layers of defense to reduce trust required in
trusted hardware.%

Lastly, Fabric (see below) has been extended~\cite{BrCaKaSo2018} to execute chaincode 
on private data using SGX.
Similarly to \pdos{}, only the contract code runs  
in the chaincode enclave, except that \pdos{} additionally make use of a 
contract interpreter. 
Differently from \pdos{}, it uses a ledger enclave to maintain (hashes of) the ledger state. 
Although it allows to verify the latest ledger state (assuming final consensus), 
it has to implement part of the Fabric peer and manage a possibly large state. 
\pdos{} instead are agnostic to the distributed ledger. Additional guarantees 
(e.g., last state verification, commitments, etc.) can be provided as required by 
leveraging features of the used distributed ledger (e.g., final consensus).

\paragraph{Distributed ledger technologies.}
Hyperledger Sawtooth~\cite{Intel2018} and Hyperledger Fabric~\cite{IBM2018, Cachin2016} are recent distributed ledger platforms.
The former is made to scale to a large number of participants.
Also, it enables customized applications and \smartcontract{}s through the concept of \emph{transaction families}, which abstracts the ledger.
The latter is designed for smaller sets of known participants and can run arbitrary \smartcontract{}s  called Chaincodes.
Also, it provides privacy and confidentiality for transactions through its \emph{channel}\footnote{Readers should be warned that the definition of \emph{channel} is counterintuitive and varies across projects.} concept, which defines a private network for a subset of the members to make transactions.
\pdo{}s can be easily built on Sawtooth by implementing custom transaction families. Also, they are able to provide stronger privacy and confidentiality guarantees with respect to Fabric.

Coco~\cite{Microsoft2018} is a framework for high-throughput and confidential blockchains for the enterprise.
Differently from \pdo{}s (and similarly to Fabric), it is designed for a known and approved set of participants.
Also, it implements leader-based consensus protocols, though its support for pluggable consensus hints that it may support decentralized protocols as Sawtooth does.
Similarly to \pdo{}s, it is based on SGX for security and performance on trusted hardware.
However, SGX is used to secure most of Coco's functionality, including core and \smartcontract{}s.
\pdo{}s instead limit SGX usage to protect the execution and data of \smartcontract{}s and do not require SGX-capable ledger validators.

\paragraph{Programming model.}
From a programming perspective, \pdo{}s can be seen as the secure extension of the abstract data objects introduced in \cite{Liskov1974}.
The extension is two-fold: (i) policies strongly bound to the data, (ii) data and policy integrity and confidentiality.
In addition, the policies defined to access and process the data can be designed to free the user from any data-related representation, or to prevent the user from using implementation details, thereby abstracting it.

\paragraph{Additional related work.}
ELI~\cite{kaptchuk2017giving} proposes a technique based on an abstract distributed ledger to implement stateful functionalities for otherwise stateless enclaves. \pdo{}s use a ledger to address a similar problem for state management. 

Town Crier~\cite{Zhang2016} is a system designed to augment \smartcontract{}s  with authenticated data feeds.
It intermediates between a user \smartcontract{} on a blockchain and a trusted data source (i.e., HTTPS-enabled web services) to deliver source-authenticated~data.
\pdo{}s can be extended to use Town Crier in future applications.

\section{Model}
\label{sec:model}

We outline the functionality and the security properties required by \pdos{}.

\subsection{Functionality}
We assume the existence of a \distributedledger{}, whose validators receive and verify transactions and answer queries about accepted transactions and stored state.
\distributedledger{} protocols such as consensus, gossip, node authentication, etc. are out of scope.

We assume the existence of SGX-enabled cloud service providers, which can provide on-demand hardware-protected execution of contracts.
In addition, we assume such enclave services can contact the manufacturer (in our case, the Intel Attestation Service) for the verification of hardware-based attestations.

\subsection{Security}
We assume that the \distributedledger{} is available and reliable. So valid transactions can be submitted and are accepted by validators, which make the data available to other parties. 
Clearly, the security properties of \pdo{}s inherit the assumptions of the underlying distributed ledger (e.g., whether a majority or two-thirds of the validators must be honest).%

We assume a computationally-bounded adversary as we rely on cryptographic primitives such as digital signatures and cryptographic hashes.
The adversary can have access to the ledger's state and tries to link transactions, contracts and data to specific users.
Also, the adversary can have access to any out-of-enclave data.

The Intel Attestation Service is assumed to be trusted. In particular, IAS correctly and completely reports whether an SGX-based  attestation was issued with cryptographic material that belongs to SGX-enabled hardware.%

Although SGX enclaves increase the cost of attacks, we assume that an adversary with enough resources can subvert an enclave. However, we assume that it is difficult for such an adversary to mount a scalable attack on the SGX platforms used in the \pdo{}'s distributed setting.%

We distinguish between attacks that compromise data confidentiality and others that target integrity. In the former case, the data of all contracts executed by the compromised enclave is potentially exposed, though it maintains its correctnetss. 
In the latter case, the correctness of the data itself can be additionally compromised. 

To mitigate these attacks, it is assumed that users are able to take action based on informed risk to calibrate the level of decentralization (e.g., how many enclaves are involved, what organization they belong to, where they are geographically located, etc.) for the provisioning of cryptographic material and the execution of contracts.

We assume that the contract interpreter and the contracts are correct. 
Specifically, the contract interpreter executes the contracts as they are intended to. Also, neither the interpreter nor the contract leak secret data.
Although the correctness property is critical for the security of \pdos{}, we leave the burden of guaranteeing it  to orthogonal research~work.

\section{Overview}
\label{sec:overview}
\label{sec:design}

\pdo{}s (\reffig{fig:pdoconcept}) allow mutually-untrusted individuals and organizations to access and update private data through pre-agreed policies. In particular, a \pdo{} implements these policies in the \smartcontract{} code ($\contractcode{}$) and binds them to the data using Intel SGX. 
Such data constitutes the \smartcontract{} state ($\contractstate{}$).
Any state update is (and can only be) the result of a \smartcontract{} method invocation. 

\begin{figure}[h]
\includegraphics[width=0.7\linewidth]{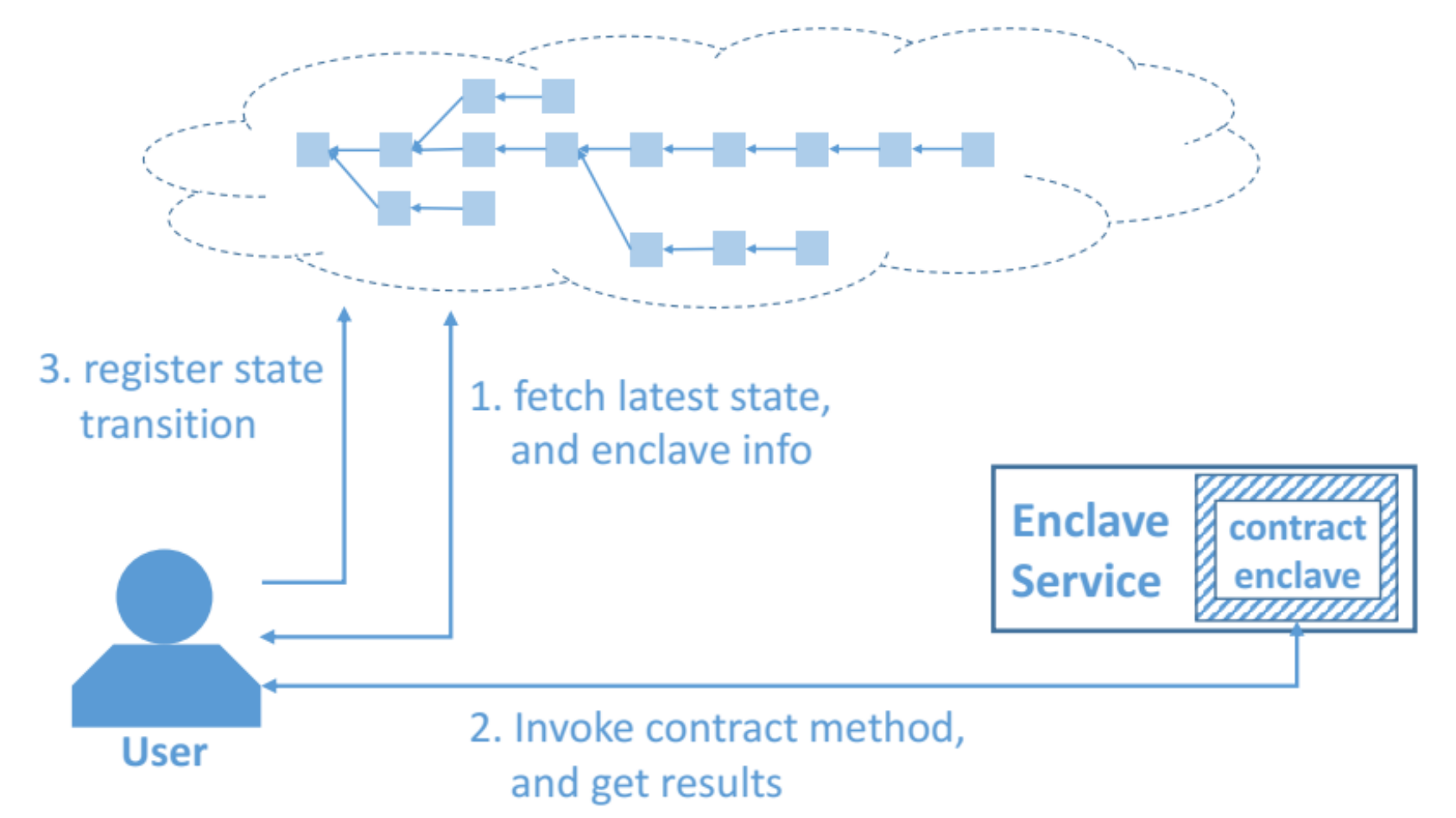}
\caption{\pdo{}'s contract method invocation.}
\label{fig:contractmethodinvocation}
\end{figure}

Method invocations (\reffig{fig:contractmethodinvocation}) are triggered by the user and performed by the  \smartcontract{} interpreter ($\contractinterpreter{}$). 
In particular, the user grabs the most recent data from the ledger and requests an invocation to the Enclave Service (\enclaveservice{}) who hosts the SGX enclave. $\contractinterpreter{}$ executes \smartcontract{} code over the input \smartcontract{} state, possibly resulting in a state transition and thus in a new state. This procedure is performed inside an SGX enclave ($\enclave{}$), which protects integrity and confidentiality of \smartcontract{} code and data. \smartcontract{} state and (optionally) code are stored in encrypted form while outside the enclave's trust boundary. 
The final results are returned to the user. The user eventually submits a transaction to the ledger  about the new state which is verified and logged for validity and later audits.

\begin{figure}[h]
\includegraphics[width=0.7\linewidth]{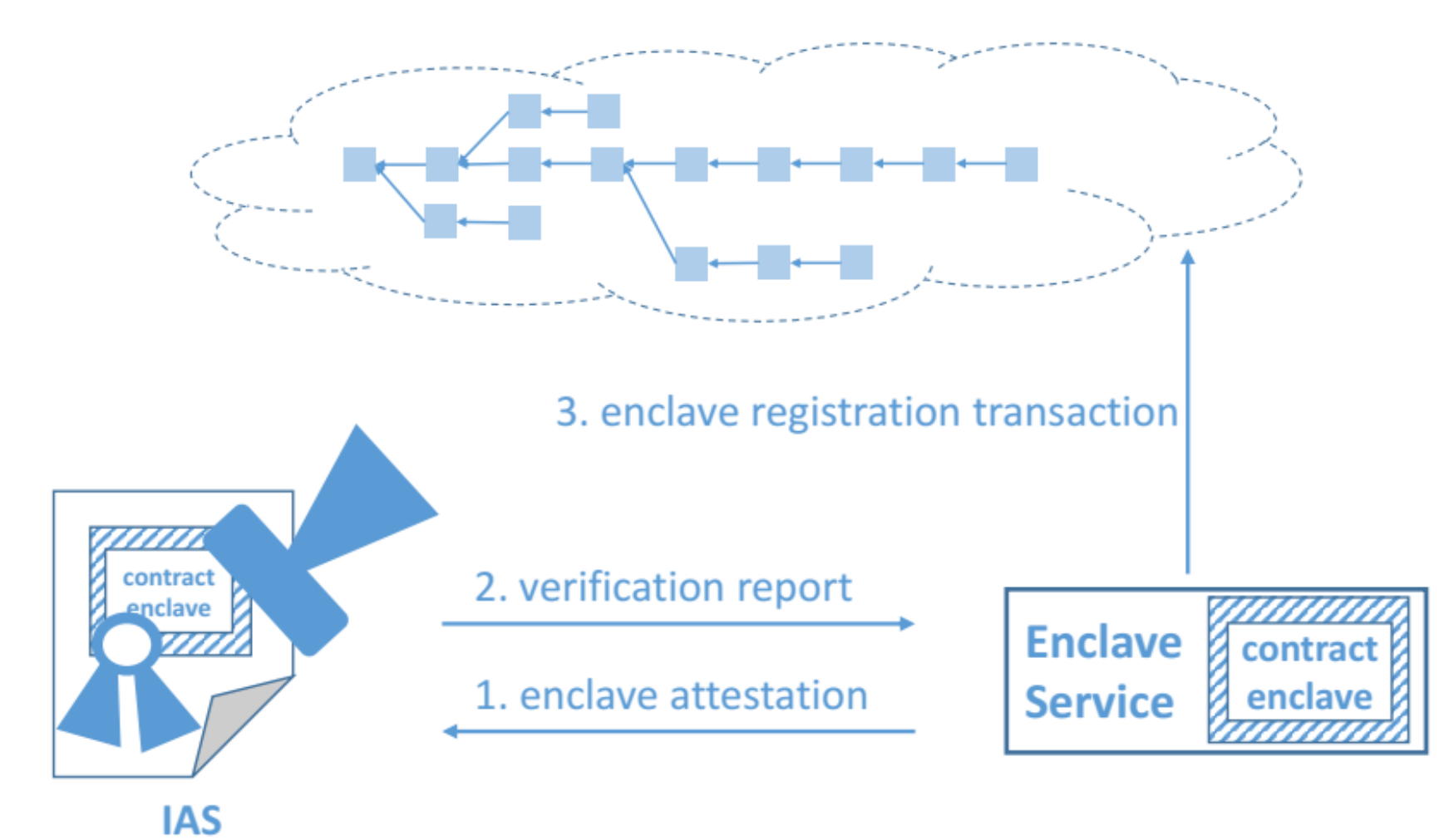}
\caption{\pdo{}'s contract enclave registration sequence. }
\label{fig:enclaveregistration}
\end{figure}

The \enclaveservice{} registers~(\reffig{fig:enclaveregistration}) the contract enclave $\contractenclave{}$ on the \distributedledger{}. The enclave is provided in the form of a \smartcontract{}-execution-as-a-service business model. 
The registration includes an enclave attestation verification step with the Intel Attestation Service (\ias{}), which verifies the enclave's hardware root-of-trust. 
The \enclaveservice{} then submits the registration transaction to the ledger together with the enclave's verification report.

\begin{figure}[h]
\includegraphics[width=0.7\linewidth]{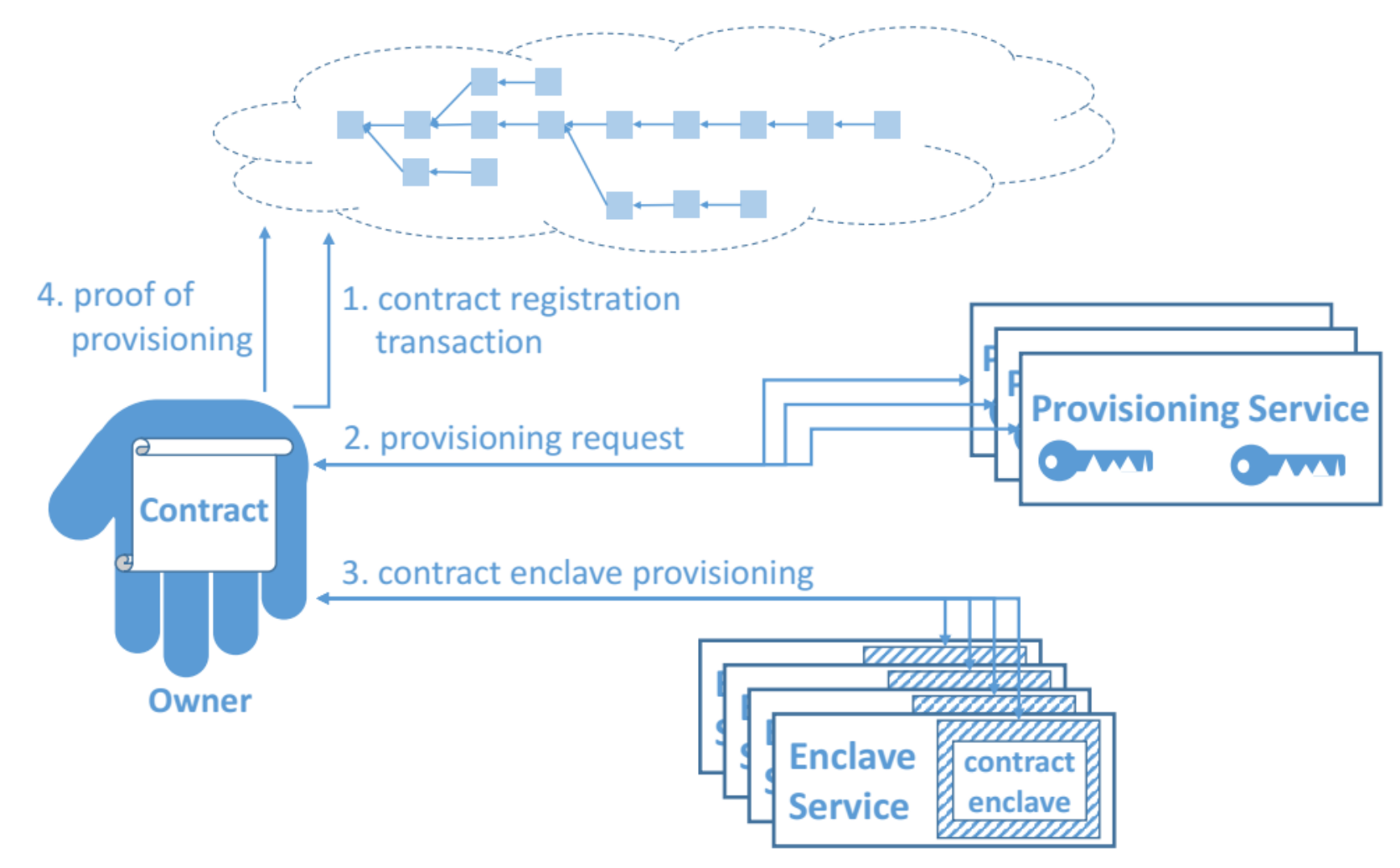}
\caption{\pdo{}s phases for contract registration and contract enclave provisioning. }
\label{fig:contractregistrationandprovisioning}
\end{figure}

Although the \smartcontract{} execution occurs off-chain, the \smartcontract{} is  registered on the \distributedledger{} by the owner, who also authorizes its execution at known \enclaveservice{}s~(\reffig{fig:contractregistrationandprovisioning}). 
The registration happens on the \distributedledger{}, while the execution authorization is realized through key provisioning. The latter is achieved by means of a set of Provisioning Services (\provisioningservice{}) which provide cryptographic material for state encryption to a registered enclave. 
The chosen \provisioningservice{}s and the actually provisioned enclaves are finally exposed on the \distributedledger{} in the contract registration transaction.

The \distributedledger{} ensures that there is a single authoritative instance and log of a \pdo{}. 
Proofs (i.e., cryptographic signatures) of the operations and identities of the involved parties are exposed to enable verification and trust establishment by end-users. 
Additionally, the \distributedledger{} helps guarantee atomicity of updates across interacting \pdo{}s and validates dependencies between transactions (if any).  
\pdo{} state transitions are not trustworthy until the associated transaction is submitted to, and accepted by, the \distributedledger{}.

\subsection{Distributed Ledger}
\label{sec:distributedledger}
\pdo{}s are designed to work jointly with a \distributedledger{}. 
The \distributedledger{} is necessary to keep track of committed transactions, dependencies, registered/updated/revoked enclaves and contracts.
Also, the \distributedledger{} validators enforce policies. For example,  on enclave registration, they check that the enclave's attestation has been verified by the Intel Attestation Service (\ias{}).

Although \pdo{}s are currently built on Hyperledger Sawtooth~\cite{Intel2018}, in principle they do not require a dedicated \distributedledger{}. 
The reason is that several \distributedledger{}s allow customization of the logic used by the validators to validate and accept transactions. 
Hence, as long as such logic is expressive enough to address the requirements of \pdo{}s, the ledger and \pdo{}s can be seen as independent components.%

Our Sawtooth-based implementation supports \pdo{}s by means of three custom transaction families. A \emph{transaction family} is the Sawtooth abstraction which allows ledger developers to implement custom rules to modify the ledger's state. Submitted transactions have to abide by such rules in order to pass the validation test and be appended to the~ledger. 

\subsubsection{Contract Registry}
\label{sec:contractregistry}
This transaction family implements operations and rules for managing contracts. 
In particular, it enables a contract owner to register and revoke a contract, and to add/remove enclaves that can run it. 
Also, the contract owner uses this registry to specify the list of provisioning services (\provisioningservice{}s) for a contract.

\subsubsection{Enclave Registry}
\label{sec:enclaveregistry}
This transaction family implements operations and rules for managing enclaves. 
The available operations include registering, updating and revoking enclaves. 
Enclave services (\enclaveservice{}s) have to comply with rules such as providing an enclave owner (who can update or revoke the enclave), the enclave's public keys (see \refsec{sec:contractenclave}) and the \ias{}-verified enclave attestation. 
This information is validated in order to accept the submitted transactions.

\subsubsection{Coordination and Commit Log (CCL)}
\label{sec:ccl}
The CCL is one the unique charactestics of \pdo{}s as it separates transaction ordering and semantics, similarly to the Global Synchronization Log (GSL)~\cite{DigitalAsset2018}. 
The CCL transaction family allows the expression of dependencies between state update transactions. 
Similarly to the GSL, the state update logic defined inside contracts is off-chain, while the contract identifiers are on-chain and used to build a dependency graph.  
The validators then enforce these dependencies by rejecting transactions whose dependencies are not committed.

These dependencies enable the coordination of the updates to one or multiple objects. 
In particular, they allow the enforcement of the serial progression of the contract state. 
Also, they allow the specification and enforcement of the progression of a contract state conditional to the progression of other contract states, while enabling maximal concurrency among interacting contracts and ledger interactions.
The validity of a progression is based on the digital signature of a contract enclave (see \refsec{sec:contractenclave}).

\subsection{Provisioning Services}
\label{sec:provisioningservice}

\provisioningservice{}s answer the contract owner ($\contractowner{}$)'s requests to provision secrets to $\contractenclave{}$. The secrets are eventually used to encrypt the state $\contractstate{}$. 
Most importantly, only a provisioned $\contractenclave{}$ has access to all secrets and so is able to decrypt the $\contractstate{}$.

The provisioning protocol is intermediated by the contract owner (\reffig{fig:contractregistrationandprovisioning}). 
The \provisioningservice{}s participate in the protocol as long as they can successfully verify the registration of contract and enclaves on the \distributedledger{}. 
The protocol can be implemented on a generic interface, which allows to customize it according to a specific application/deployment scenario. 
The currently implemented protocol let the \provisioningservice{}s generate secrets, and it guarantees security as long as one of the participating \provisioningservice{}s is honest.

The purpose of using \provisioningservice{}s is to make $\contractstate{}$ encryption (i) enclave-independent and (ii) \enclaveservice{}-independent. 
The first property is important to ensure that contract enclaves are stateless with respect to contracts (i.e., they only need a contract and the contract state to perform a method invocation). This prevents the problem of how to move the state between two enclaves running on separate platforms. 
The second property is necessary because an \enclaveservice{} is considered untrusted by other parties. Once the enclave terminates and transactions are committed on the ledger, the \enclaveservice{} cannot be trusted to persist critical enclave's data (e.g., sealed storage and encrypted state). This would open it  to trivial albeit severe denial of service attacks. 

Notice that the DoS issue is strictly related to $\contractstate{}$, and not to any enclave's data in sealed storage. The enclave in fact has some private cryptographic material, which is sealed and should be persisted for later invocations. However, the unavailability of this data only makes the associated enclave's execution environment unavailable, since the enclave cannot recover its private cryptographic material. In particular, this problem does not prevent the use of other enclaves, possibly at other \enclaveservice{}s.

\subsection{Enclave Hosting Service}
\label{sec:enclavehostingservice}
An Enclave Service (\enclaveservice{}) is an entity that owns SGX-enabled platforms, and provides generic SGX-based Contract Enclaves as a service (e.g., a cloud provider). 
However, ownership of the SGX-enabled hardware is not sufficient for \pdo{}s. 
\enclaveservice{}s additionally have to comply with the enclave registration/update/revocation requirements on the \distributedledger{}, and to prove that such enclaves run the correct software. 
The latter is achieved by verifying SGX enclaves with \ias{}, while the former is achieved by submitting to the \distributedledger{} nodes valid transactions, which have to include \ias{} signed verification reports.

\subsection{Contract Enclave}
\label{sec:contractenclave}
Contract Enclaves ($\contractenclave{}$) are SGX enclaves which include a contract interpreter ($\contractinterpreter{}$) and thus can run contracts. 
For the sake of making them general, they do not include any contract or data. 
Hence, one $\contractenclave{}$ can handle many \pdo{} instances, i.e., $\contract{}, \contractstate{}$ pairs. 
In this sense, an \enclaveservice{} provides a $\contractenclave{}$-as-a-service.

During its lifetime, a $\contractenclave{}$ is registered into/revoked from the enclave registry (\refsec{sec:enclaveregistry}), added to/removed from the list of enclaves authorized to execute a specific contract $\contract{}$ in the contract registry  (\refsec{sec:contractregistry}). Also, it generates execution results, which are described in transactions submitted to the CCL (\refsec{sec:ccl}) for validation. 

An \enclaveservice{} leverages \ias{} to let others (i.e., \distributedledger{} validators) establish trust in a $\contractenclave{}$. Then, $\contractenclave{}$-generated and attested key pairs are used to preserve trust and extend it to the enclave's signed statements. In particular, the $\contractenclave{}$ SGX attestation is issued and submitted to the validators as a proof of correct execution. The attestation covers (and thus extends trust to) two $\contractenclave{}$ public keys, namely: a public verification key, which is used to check the enclave's signatures and to limit the interactions with \ias{}; and a public encryption key, which is used to provision secrets to the enclave. The two public/private key pairs are generated by the enclave, so they are bound to a specific enclave instance. The public keys are attested, registered and retrievable from the \distributedledger{}, while their private counterparts are saved by the enclave in sealed storage.

Although the $\contractenclave{}$ is responsible for the confidentiality of $\contractstate{}$, the encryption key is not bound to the enclave itself, 
but rather securely provisioned to it on a per-contract and per-contract-owner basis. 
$\contractenclave{}$ verifies the  signatures of the \provisioningservice{}s over the provisioned secrets. Eventually, it signs a proof of provisioning, which includes the \provisioningservice{}s and the (encrypted) secrets.
Ultimately, it is the contract owner that adds such provisioned enclave to the list of enclaves (on the \distributedledger{}) which are authorized to run the contract.

\subsubsection{Contract Interpreter}

Similarly to the Ethereum EVM~\cite{Wood2014},  
the contract interpreter $\contractinterpreter{}$ executes a contract code $\contractcode{}$ registered on the \distributedledger{} by a contract owner $\contractowner{}$. 
Most importantly, $\contractinterpreter{}$ is not bound to a specific $\contract{}$, so $\contractenclave{}$ can run any contract written in the interpreter's~language.

Since $\contractinterpreter{}$ runs in an SGX enclave, there are some features that are desirable in an interpreter's implementation. 
\begin{enumerate*}[label=(\roman*)]
\item \emph{Contract code sandboxing}. This prevents (possibly) malicious contracts implemented by (possibly) untrusted third-parties to tamper with the rest of the enclave and the \enclaveservice{} platform.
\item \emph{Small code footprint/TCB}. This arguably reduces the complexity of the code, thereby simplifying code audit.
\item \emph{Unique contract representation}. This makes a contract easily verifiable by the participants. Also, all participants can agree on the same copy, which can be identified on the \distributedledger{} with one cryptographic hash, thereby avoiding multiple documents due to non-deterministic compilations. 
Finally, it easily enables the authentication and execution of the agreed copy within the enclave. 
\item \emph{Efficient contract state management}. This allows a contract to process the state without requiring to load it upfront inside the enclave. 
\end{enumerate*}

\pdo{}s leverage a small abstraction inside $\contractenclave{}$ for generic contract interpreters. 
The current default implementation relies on \emph{\tinyscheme{}}~\cite{Souflis2005}, which is based on the {\emph{\scheme{}}}~\cite{Sussman1975,Sussman1998} functional language. We plan to support additional interpreters.

\subsubsection{Contract}
A contract ($\contract{}$) defines and implements methods for accessing and updating data in the contract state. 
As parties need to agree on a $\contract{}$ before using it in a \pdo{}, it is expected that they communicate with each other offline in order to share the contract code offchain. 
So the contract must not necessarily be public. Its execution is designed to occur natively off-chain inside SGX enclaves at the chosen \enclaveservice{}s (\refsec{sec:enclavehostingservice}).

\subsubsection{Contract State}
The contract state $\contractstate{}$ indicates the (private) data which a \pdo{} associates to a specific contract. 
At $\contract{}$ runtime, the contract has access to plaintext state, which is managed by  $\contractinterpreter{}$. 
In the other circumstances, $\contractstate{}$ is encrypted while stored outside of the contract enclave.

Let us provide some details of the workflow.  
Upon the first contract method invocation, $\contractinterpreter{}$ initializes an empty state $\contractstate{}$ for $\contract{}$. 
In subsequent invocations, $\contractinterpreter{}$ reads or modifies $\contractstate{}$ according to the logic implemented by $\contract{}$. 
$\contractinterpreter{}$ can access $\contractstate{}$ in plaintext, since SGX maintains the confidentiality of the data. 
$\contractenclave{}$ decrypts (resp. encrypts) $\contractstate{}$ before (resp. after) $\contract{}$ requires data. These operations are performed using a contract-specific key, which is derived from the provisioned secrets (see \refsec{sec:provisioningservice}).

\subsection{Contract Owner}
The contract owner $\contractowner{}$ is the person/organization that creates a \pdo{} instance by registering a contract $\contract{}$ and mediating the provisioning of one or more $\contractenclave{}$s.
The registration is done by submitting a transaction on the \distributedledger{}.  
The provisioning happens by contacting the chosen \provisioningservice{}s and ensuring that $\contractenclave{}$ was provisioned. 
The proof of correct provisioning is exposed on the \distributedledger{}, thereby showing that an $\contractenclave{}$ is able to run the registered $\contract{}$.
The contract creation happens through method invocation in a $\contractenclave{}$, which (being the first invocation)  initializes an empty $\contractstate{}$.

\subsection{Users}
\label{sec:users}
Users transact  together in three phases. 
First, they send method invocation messages to contracts. User authentication can be performed by  $\contractenclave{}$ %
or by a contract according to the implemented policies. In this way  only authorized participants are able to invoke methods.
Second, they check the results returned by a $\contractenclave{}$ and delivered to them through the  \enclaveservice{}.
Third, they submit the resulting transaction to the \distributedledger{}. 

The design rationale for such protocol stems from the mutual-distrust between the parties (including the \enclaveservice{}s). 
At the \enclaveservice{}s, as the helper process for $\contractenclave{}$ is not in the TCB, it cannot be trusted to submit a transaction to the ledger or handle communication between objects. 
In a specular manner, if a user decides not to commit the results, the helper process cannot be trusted not to do so. 

\pdo{}s leverage the user's motivation to invoke a method and get results to solve this problem.  
In particular, a user acts as a \emph{channel} for communication between the enclave and the ledger (and additionally between objects). 

\pdo{}s implement the channel concept by enforcing that only the one who invokes a method can submit the resulting transaction to the \distributedledger{}. 
The channel is set up as follows: the user generates a fresh key pair at each invocation, includes the public key in the method invocation message and signs the message; after the method invocation,  $\contractenclave{}$ includes the public key in the results; finally, the \distributedledger{} validators  only accept the resulting transaction if it is also signed by the holder of the corresponding (channel's) private key. 
As a different key pair is used for each transaction commit, this prevents \distributedledger{} validators from correlating transactions, thereby protecting the user's privacy.

\section{Security Aspects of \pdo{}'s Decentralized Model}
\label{sec:securityanalysis}

A critical objective of \pdo{}s is to avoid the presence of single points of failures that can jeopardize the entire system. Once this can be assured, the next step is how to detect and remove failed end-points, such as compromised enclaves. In this section we elaborate on these key security aspects.

\subsection{Trust but verify what is on the ledger}
As participants are mutually-untrusted, all of their interactions are based on mechanisms that allow to verify each other's work. We briefly summarize the verification capabilities of the participants.

First, the \distributedledger{} is designed to distribute trust among the validators. As we mentioned however, its design is orthogonal to \pdo{}s.

Second, the contract owner entrusts \provisioningservice{}s to provision an enclave, and an \enclaveservice{} to execute it.
However, the owner later verifies the correct enclave execution, secret provisioning and contract/enclave registration transaction on the \distributedledger{}.
An \enclaveservice{} submits but verifies enclave registration/revocation transactions.
\provisioningservice{}'s verify with the \distributedledger{} that they are going to provision actual enclaves before moving forward.
Users entrust an \enclaveservice{} to execute a contract enclave, but verify the enclave's work and directly submit transactions.

Additionally, the users are able to verify the provisioning procedure.
The contract owner in fact publishes on the \distributedledger{} the identities of the selected \provisioningservice{}s, together with their provisioned encrypted secrets and a proof of correct enclave provisioning.
So users can check on the \distributedledger{} whether the \provisioningservice{}s belong to one or more possibly reputable organizations and whether enough \provisioningservice{}s have been selected.
Ultimately, this allows users to trust that the encryption key is secure and only available to the provisioned enclave.

The encryption key plays a critical role in state confidentiality but not in its integrity.
If an adversary compromises the key, then exposing the state is a trivial task.
However, the adversary cannot tamper with the integrity of the state in order to corrupt a future committed state.
The reason is that \distributedledger{} validators only accept a CCL transaction (\refsec{sec:ccl}) if an enclave signs a valid state update (i.e., a transition from the last committed state to another).
So if the adversary tampers with the state before/after the execution, the validators would be unable to accept the transaction.

\subsection{Compromised Enclaves}
\pdo{}s aim at taking advantage of SGX without ruling out the possibility that enclaves could be compromised.
As enclaves deal with confidential data, this would trivially result in the exposure of the handled data.
Since \pdo{}s natively use per-\pdo{} state encryption keys, the problem would impact only the contracts running on the compromised enclave.

The design however enables mechanisms for limiting the confidentiality breach and for maintaining integrity.
The first mechanism is to limit the explicitly provisioned enclaves and possibly to run them at reputable \enclaveservice{}s.
The mechanism enables a contract owner to make these decisions and also allows the users to take informed risk.
In fact,
the details of the \provisioningservice{}s are exposed on the \distributedledger{}.

Another mechanism is the verification of an enclave's work.
So while confidentiality is compromised, this still allows to preserve the integrity of the computation.
\pdo{}s can implement verification by letting additional enclaves re-execute a method using the same input parameters.
If their outcomes match, the output of an enclave can be used to commit the transaction.
Otherwise, the majority of matching responses (ideally $n-1$ out of $n$) can be considered as a proof of misbehavior of the remaining enclaves.
This could result for example in enclave revocation and/or state commit revocation transactions.

\section{Conclusions}
\label{sec:conclusions}

\pdo{}s introduce a new way to securely run smart contracts over private data by leveraging Intel SGX. They are designed following a decentralized trust model in order to work in a network of mutually-untrusted entities and participants. In addition, they take advantage of a distributed ledger which can maintain a single authoritative instance of the objects. Their design is ledger-agnostic, so the implementation can be adapted for different ledger architectures.

From a security perspective, \pdo{}s protect user privacy on the ledger and data confidentiality in secure enclaves. The security architecture exposes enclave hosting services and enclave provisioning services on the ledger (upon registration). This enables participants to establish trust in the enclaves based on informed risk (i.e., informing about who runs them and who provisions them). Most importantly, \pdo{}s are designed to limit confidentiality breaches, to provide mechanisms to guarantee integrity in spite of an enclave compromise, and to revoke misbehaving~enclaves.



\begin{thebibliography}{20}


\ifx \showCODEN    \undefined \def \showCODEN     #1{\unskip}     \fi
\ifx \showDOI      \undefined \def \showDOI       #1{#1}\fi
\ifx \showISBNx    \undefined \def \showISBNx     #1{\unskip}     \fi
\ifx \showISBNxiii \undefined \def \showISBNxiii  #1{\unskip}     \fi
\ifx \showISSN     \undefined \def \showISSN      #1{\unskip}     \fi
\ifx \showLCCN     \undefined \def \showLCCN      #1{\unskip}     \fi
\ifx \shownote     \undefined \def \shownote      #1{#1}          \fi
\ifx \showarticletitle \undefined \def \showarticletitle #1{#1}   \fi
\ifx \showURL      \undefined \def \showURL       {\relax}        \fi
\providecommand\bibfield[2]{#2}
\providecommand\bibinfo[2]{#2}
\providecommand\natexlab[1]{#1}
\providecommand\showeprint[2][]{arXiv:#2}

\bibitem[\protect\citeauthoryear{Brandenburger, Cachin, Kapitza, and
  Sorniotti}{Brandenburger et~al\mbox{.}}{2018}]%
        {BrCaKaSo2018}
\bibfield{author}{\bibinfo{person}{Marcus Brandenburger},
  \bibinfo{person}{Christian Cachin}, \bibinfo{person}{R{\"u}diger Kapitza},
  {and} \bibinfo{person}{Alessandro Sorniotti}.}
  \bibinfo{year}{2018}\natexlab{}.
\newblock \bibinfo{booktitle}{{\em Blockchain and Trusted Computing: Problems,
  Pitfalls, and a Solution for {Hyperledger} {Fabric}}}.
\newblock \bibinfo{type}{{T}echnical {R}eport} arXiv:1805.08541v1 [cs.DC].
  \bibinfo{institution}{arXiv.org}.
\newblock


\bibitem[\protect\citeauthoryear{Cachin}{Cachin}{2016}]%
        {Cachin2016}
\bibfield{author}{\bibinfo{person}{Christian Cachin}.}
  \bibinfo{year}{2016}\natexlab{}.
\newblock \showarticletitle{Architecture of the Hyperledger blockchain fabric}.
  In \bibinfo{booktitle}{{\em Workshop on Distributed Cryptocurrencies and
  Consensus Ledgers}}.
\newblock


\bibitem[\protect\citeauthoryear{Cecchetti, Zhang, Ji, Kosba, Juels, and
  Shi}{Cecchetti et~al\mbox{.}}{2017}]%
        {Cecchetti2017}
\bibfield{author}{\bibinfo{person}{Ethan Cecchetti}, \bibinfo{person}{Fan
  Zhang}, \bibinfo{person}{Yan Ji}, \bibinfo{person}{Ahmed Kosba},
  \bibinfo{person}{Ari Juels}, {and} \bibinfo{person}{Elaine Shi}.}
  \bibinfo{year}{2017}\natexlab{}.
\newblock \showarticletitle{Solidus: Confidential Distributed Ledger
  Transactions via PVORM}. In \bibinfo{booktitle}{{\em Proceedings of the 2017
  ACM SIGSAC Conference on Computer and Communications Security (CCS)}}.
  \bibinfo{pages}{701--717}.
\newblock
\showISBNx{978-1-4503-4946-8}


\bibitem[\protect\citeauthoryear{Cheng, Zhang, Kos, He, Hynes, Johnson, Juels,
  Miller, and Song}{Cheng et~al\mbox{.}}{2018}]%
        {cheng2018ekiden}
\bibfield{author}{\bibinfo{person}{Raymond Cheng}, \bibinfo{person}{Fan Zhang},
  \bibinfo{person}{Jernej Kos}, \bibinfo{person}{Warren He},
  \bibinfo{person}{Nicholas Hynes}, \bibinfo{person}{Noah Johnson},
  \bibinfo{person}{Ari Juels}, \bibinfo{person}{Andrew Miller}, {and}
  \bibinfo{person}{Dawn Song}.} \bibinfo{year}{2018}\natexlab{}.
\newblock \showarticletitle{Ekiden: A Platform for Confidentiality-Preserving,
  Trustworthy, and Performant Smart Contract Execution}.
\newblock \bibinfo{journal}{{\em arXiv preprint arXiv:1804.05141\/}}
  (\bibinfo{year}{2018}).
\newblock


\bibitem[\protect\citeauthoryear{Corp.}{Corp.}{2018}]%
        {Privatedataobjectrepo2018}
\bibfield{author}{\bibinfo{person}{Intel Corp.}}
  \bibinfo{year}{2018}\natexlab{}.
\newblock \showarticletitle{Private Data Objects}.
\newblock \bibinfo{journal}{{\em
  \customurl{https://github.com/hyperledger-labs/private-data-objects}\/}}
  (\bibinfo{year}{2018}).
\newblock


\bibitem[\protect\citeauthoryear{DigitalAsset}{DigitalAsset}{2018}]%
        {DigitalAsset2018}
\bibfield{author}{\bibinfo{person}{DigitalAsset}.}
  \bibinfo{year}{2018}\natexlab{}.
\newblock \showarticletitle{The Global Synchronization Log}.
\newblock \bibinfo{journal}{{\em
  \customurl{http://hub.digitalasset.com/hubfs/Documents/TheGlobalSynchronizationLog.pdf}\/}}
  (\bibinfo{year}{2018}).
\newblock


\bibitem[\protect\citeauthoryear{Foundation}{Foundation}{2018}]%
        {Hyperledgerlabs2018}
\bibfield{author}{\bibinfo{person}{Linux Foundation}.}
  \bibinfo{year}{2018}\natexlab{}.
\newblock \showarticletitle{Hyperledger Labs}.
\newblock \bibinfo{journal}{{\em
  \customurl{https://hyperledger-labs.github.io/}\/}} (\bibinfo{year}{2018}).
\newblock


\bibitem[\protect\citeauthoryear{IBM}{IBM}{2018}]%
        {IBM2018}
\bibfield{author}{\bibinfo{person}{IBM}.} \bibinfo{year}{2018}\natexlab{}.
\newblock \showarticletitle{Hyperledger Fabric}.
\newblock \bibinfo{journal}{{\em
  \customurl{https://www.hyperledger.org/projects/fabric}\/}}
  (\bibinfo{year}{2018}).
\newblock


\bibitem[\protect\citeauthoryear{Intel}{Intel}{2018}]%
        {Intel2018}
\bibfield{author}{\bibinfo{person}{Intel}.} \bibinfo{year}{2018}\natexlab{}.
\newblock \showarticletitle{Hyperledger Sawtooth}.
\newblock \bibinfo{journal}{{\em
  \customurl{https://www.hyperledger.org/projects/sawtooth}\/}}
  (\bibinfo{year}{2018}).
\newblock


\bibitem[\protect\citeauthoryear{Kaptchuk, Miers, and Green}{Kaptchuk
  et~al\mbox{.}}{2017}]%
        {kaptchuk2017giving}
\bibfield{author}{\bibinfo{person}{Gabriel Kaptchuk}, \bibinfo{person}{Ian
  Miers}, {and} \bibinfo{person}{Matthew Green}.}
  \bibinfo{year}{2017}\natexlab{}.
\newblock \bibinfo{booktitle}{{\em Giving State to the Stateless: Augmenting
  Trustworthy Computation with Ledgers}}.
\newblock \bibinfo{type}{{T}echnical {R}eport}.
  \bibinfo{institution}{Cryptology ePrint Archive, Report 2017/201, 2017.
  https://eprint. iacr. org/2017/201}.
\newblock


\bibitem[\protect\citeauthoryear{Kosba, Miller, Shi, Wen, and
  Papamanthou}{Kosba et~al\mbox{.}}{2016}]%
        {Kosba2016}
\bibfield{author}{\bibinfo{person}{Ahmed Kosba}, \bibinfo{person}{Andrew
  Miller}, \bibinfo{person}{Elaine Shi}, \bibinfo{person}{Zikai Wen}, {and}
  \bibinfo{person}{Charalampos Papamanthou}.} \bibinfo{year}{2016}\natexlab{}.
\newblock \showarticletitle{Hawk: The blockchain model of cryptography and
  privacy-preserving smart contracts}. In \bibinfo{booktitle}{{\em Security and
  Privacy (SP), 2016 IEEE Symposium on}}. IEEE, \bibinfo{pages}{839--858}.
\newblock


\bibitem[\protect\citeauthoryear{Liskov and Zilles}{Liskov and Zilles}{1974}]%
        {Liskov1974}
\bibfield{author}{\bibinfo{person}{Barbara Liskov} {and}
  \bibinfo{person}{Stephen Zilles}.} \bibinfo{year}{1974}\natexlab{}.
\newblock \showarticletitle{Programming with Abstract Data Types}. In
  \bibinfo{booktitle}{{\em Proceedings of the ACM SIGPLAN Symposium on Very
  High Level Languages}}. \bibinfo{pages}{50--59}.
\newblock


\bibitem[\protect\citeauthoryear{Microsoft}{Microsoft}{2018}]%
        {Microsoft2018}
\bibfield{author}{\bibinfo{person}{Microsoft}.}
  \bibinfo{year}{2018}\natexlab{}.
\newblock \showarticletitle{The Coco Framework}.
\newblock \bibinfo{journal}{{\em
  \customurl{https://github.com/Azure/coco-framework/blob/master/docs/Coco\%20Framework\%20whitepaper.pdf}\/}}
  (\bibinfo{year}{2018}).
\newblock


\bibitem[\protect\citeauthoryear{Sasson, Chiesa, Garman, Green, Miers, Tromer,
  and Virza}{Sasson et~al\mbox{.}}{2014}]%
        {Sasson2014}
\bibfield{author}{\bibinfo{person}{Eli~Ben Sasson}, \bibinfo{person}{Alessandro
  Chiesa}, \bibinfo{person}{Christina Garman}, \bibinfo{person}{Matthew Green},
  \bibinfo{person}{Ian Miers}, \bibinfo{person}{Eran Tromer}, {and}
  \bibinfo{person}{Madars Virza}.} \bibinfo{year}{2014}\natexlab{}.
\newblock \showarticletitle{Zerocash: Decentralized anonymous payments from
  bitcoin}. In \bibinfo{booktitle}{{\em Security and Privacy (SP), 2014 IEEE
  Symposium on}}. IEEE, \bibinfo{pages}{459--474}.
\newblock


\bibitem[\protect\citeauthoryear{Souflis and Shapiro}{Souflis and
  Shapiro}{2005}]%
        {Souflis2005}
\bibfield{author}{\bibinfo{person}{Dimitrios Souflis} {and} \bibinfo{person}{J
  Shapiro}.} \bibinfo{year}{2005}\natexlab{}.
\newblock \showarticletitle{TinyScheme}.
\newblock \bibinfo{journal}{{\em
  \customurl{http://tinyscheme.sourceforge.net}\/}} (\bibinfo{year}{2005}).
\newblock


\bibitem[\protect\citeauthoryear{Sussman and Steele}{Sussman and
  Steele}{1998}]%
        {Sussman1998}
\bibfield{author}{\bibinfo{person}{Gerald~Jay Sussman} {and}
  \bibinfo{person}{Guy~L. Steele, Jr.}} \bibinfo{year}{1998}\natexlab{}.
\newblock \showarticletitle{Scheme: A Interpreter for Extended Lambda
  Calculus}.
\newblock \bibinfo{journal}{{\em Higher Order Symbol. Comput.\/}}
  \bibinfo{volume}{11}, \bibinfo{number}{4} (\bibinfo{date}{Dec.}
  \bibinfo{year}{1998}), \bibinfo{pages}{405--439}.
\newblock
\showISSN{1388-3690}


\bibitem[\protect\citeauthoryear{Sussman and Steele~Jr}{Sussman and
  Steele~Jr}{1975}]%
        {Sussman1975}
\bibfield{author}{\bibinfo{person}{Gerald~J Sussman} {and}
  \bibinfo{person}{Guy~L Steele~Jr}.} \bibinfo{year}{1975}\natexlab{}.
\newblock \bibinfo{booktitle}{{\em SCHEME: An Interpreter for Extended Lambda
  Calculus}}.
\newblock \bibinfo{type}{{T}echnical {R}eport} AI Memo n.349.
  \bibinfo{institution}{MIT Artificial Intelligence Laboratory}.
\newblock


\bibitem[\protect\citeauthoryear{Wood}{Wood}{2014}]%
        {Wood2014}
\bibfield{author}{\bibinfo{person}{Gavin Wood}.}
  \bibinfo{year}{2014}\natexlab{}.
\newblock \showarticletitle{Ethereum: A secure decentralised generalised
  transaction ledger}.
\newblock \bibinfo{journal}{{\em Ethereum Project Yellow Paper\/}}
  \bibinfo{volume}{151} (\bibinfo{year}{2014}), \bibinfo{pages}{1--32}.
\newblock


\bibitem[\protect\citeauthoryear{Zhang, Cecchetti, Croman, Juels, and
  Shi}{Zhang et~al\mbox{.}}{2016}]%
        {Zhang2016}
\bibfield{author}{\bibinfo{person}{Fan Zhang}, \bibinfo{person}{Ethan
  Cecchetti}, \bibinfo{person}{Kyle Croman}, \bibinfo{person}{Ari Juels}, {and}
  \bibinfo{person}{Elaine Shi}.} \bibinfo{year}{2016}\natexlab{}.
\newblock \showarticletitle{Town Crier: An Authenticated Data Feed for Smart
  Contracts}. In \bibinfo{booktitle}{{\em Proceedings of the 2016 ACM SIGSAC
  Conference on Computer and Communications Security (CCS)}}.
  \bibinfo{pages}{270--282}.
\newblock
\showISBNx{978-1-4503-4139-4}


\bibitem[\protect\citeauthoryear{Zyskind, Nathan, and Pentland}{Zyskind
  et~al\mbox{.}}{2015}]%
        {Zyskind2015}
\bibfield{author}{\bibinfo{person}{Guy Zyskind}, \bibinfo{person}{Oz Nathan},
  {and} \bibinfo{person}{Alex Pentland}.} \bibinfo{year}{2015}\natexlab{}.
\newblock \showarticletitle{Enigma: Decentralized computation platform with
  guaranteed privacy}.
\newblock \bibinfo{journal}{{\em arXiv preprint arXiv:1506.03471\/}}
  (\bibinfo{year}{2015}).
\newblock


\end{thebibliography}
\end{document}